\documentclass[preprint,acs]{revtex4}
\usepackage{graphicx}
\usepackage{dcolumn}
\usepackage{bm}

\begin{document}

\begin{titlepage}

\title{Adsorption of gas molecules on graphene nanoribbons and its
implication for nano-scale molecule sensor}
%: a first principles study}

\author{Bing Huang$^1$, Zuanyi Li$^1$, Zhirong Liu$^2$, Gang Zhou$^1$, Shaogang Hao$^1$, Jian Wu$^1$,
Bing-Lin Gu$^1$ and Wenhui Duan$^1$\footnote{Author to whom
correspondence should be addressed. E-mail address:
dwh@phys.tsinghua.edu.cn}}
\address{$^1$Department of Physics, Tsinghua University, Beijing
100084, People's Republic of China \\ $^2$College of Chemistry and
Molecular Engineering, Peking University, Beijing 100871, People's
Republic of China}
\date{\today}

\begin{abstract}

We have studied the adsorption of gas molecules (CO, NO, NO$_{2}$,
O$_{2}$, N$_{2}$, CO$_{2}$, and NH$_{3}$) on graphene nanoribbons
(GNRs) using first principles methods. The adsorption geometries,
adsorption energies, charge transfer, and electronic band
structures are obtained. We find that the electronic and transport
properties of the GNR with armchair-shaped edges are sensitive to
the adsorption of NH$_{3}$ and the system exhibits \emph{n}-type
semiconducting behavior after NH$_{3}$ adsorption. Other gas
molecules have little effect on modifying the conductance of GNRs.
Quantum transport calculations further indicate that NH$_{3}$
molecules can be detected out of these gas molecules by GNR based
sensor.

\end{abstract}

\maketitle

\draft

\vspace{2mm}

\end{titlepage}

\section{Introduction}

Sensing gas molecules is critical to environmental monitoring,
control of chemical processes, space missions, and agricultural
and medical applications\cite{MRS}. Solid-state gas sensors are
renowned for their high sensitivity which have made them
ubiquitous in the world\cite{Moseley, Capone}. In the past few
years, a new generation of gas sensors have been demonstrated
using carbon nanotubes (CNTs) and semiconductor
nanowires\cite{Kong, Collins, Qi, Valentini, Novak, Li, Zhang,
Chopra}. It was reported that semiconducting CNTs could be used to
detect small concentration of NH$_{3}$, NO$_{2}$, and O$_{2}$ with
high sensitivity by measuring changes of the CNTs conductance upon
exposure to the gases at room temperature\cite{Kong, Collins, Qi,
Valentini, Chopra}.

Graphene, a single atomic layer of graphite, has been successfully
produced in experiments\cite{K. S. Novoselov, C. Berger}, which
have resulted in intensive investigations on graphene-based
structures because of fundamental physics interests and promising
applications\cite{A. K. Geim} (\emph{e.g.}, gas
sensor\cite{F.Schedin}). Importantly, graphene can be patterned
via standard lithographic techniques into quasi-one-dimension
materials\cite{C. Berger, Kim, Chen}, graphene nanoribbons (GNRs),
which have many properties similar to CNTs, such as energy gap
dependence of widths and crystallographic orientation\cite{Qimin,
Louie}. Different from CNTs, however, GNRs present long and
reactive edges which make GNRs not only notably more accessible to
doping\cite{Qimin, Bing, Bing2} and chemical
modification\cite{Wang, Hod, Hao, Ferrari}, but also more
susceptible to structural defects and impurites\cite{Bing2}.

Theoretical studies of gas molecular adsorption on the graphene
surface have been reported recently\cite{Wehling, O.Leenaerts},
which showed that NO$_{2}$, H$_{2}$O, NH$_{3}$, CO and NO molecules
are physically adsorbed on the pristine graphene: NH$_{3}$ and CO
molecules will act as donors while H$_{2}$O and NO$_{2}$ will act as
acceptors, which are consistent with previous
experiment\cite{F.Schedin}. Compared with graphene, GNRs are
advantageous in small volume and free reactive edges. In
experiments, the edges of GNRs are not well controlled\cite{Chen,
Kim} and it is hard to obtain fully saturated edges without any
dangling bond (DB) defects. It is well known that DB defects around
the vacancy sites or at the tips play a very important role in CNTs
gas sensors because they are very chemically reactive\cite{Snow,
Andzelm, Vladimir}. Similar to CNTs, when there are DB defects at
GNRs edges, covalent attachment of chemical groups and molecules
also significantly influences their electronic properties\cite{Wang,
Hao, Ferrari}. Therefore, it is very interesting and important to
study the feasibility of using GNRs as gas sensors.

In this article, using density functional theory (DFT)
calculations, we study the adsorption of gas molecules (CO, NO,
NO$_{2}$, O$_{2}$, N$_{2}$, CO$_{2}$ and NH$_{3}$) around the
sites of DB defects on armchair GNRs (AGNRs, having
armchair-shaped edges) and explore the feasibility of using AGNRs
as gas sensors. Following conventional notation\cite{defination},
a GNR is specified by the numbers ($n$) of dimer lines and zigzag
chains along the ribbon forming the width, for the AGNR and zigzag
GNRs (ZGNRs, having zigzag-shaped edges), respectively. For
example, the structure in Fig. 1a is referred as a 10-AGNR (i.e.,
$n=10$). Previous works show that all AGNRs are semiconductor
while ZGNRs are metal\cite{defination,Qimin}. We focus on
semiconducting AGNRs instead of metallic ZGNRs, since it is
expected that gas molecule adsorption will have a much smaller
effect on modifying the electronic properties of (metallic) ZGNRs.
It is found that although all gas molecules can influence the
electronic structure of AGNRs, only NH$_{3}$ molecule adsorption
can modify the conductance of AGNRs remarkably by acting as donors
while other gas molecules have little effect on conductance. This
property can be utilized to detect NH$_{3}$ out of other common
gases, which is requisite and significant in industrial, medical,
and living environments\cite{Takao}.

\section{Calculation Method and Model}
Our electronic structure calculations were performed using the DFT
in the spin-polarized generalized gradient approximation (GGA)
with PW91 functional for the exchange and correlation effects of
the electrons, as implemented in Vienna \emph{Ab initio}
Simulation Package (VASP)\cite{VASP}. The electron-ion interaction
was described by the ultrasoft pseudopentials and the energy
cutoff was set to be 400 eV. Structural optimization was carried
out on all systems until the residual forces on all ions were
converged to 0.01 eV/\AA. The quantum transport calculations were
performed using the ATK 2.3 package\cite{Taylor}, which implements
DFT-based real-space, nonequilibrium Green's function formalism.
The mesh cutoff is chosen as 150 Ry to achieve the balance between
calculation efficiency and accuracy.

%fig01
\begin{figure}[tbp]
\includegraphics[width=8.5cm]{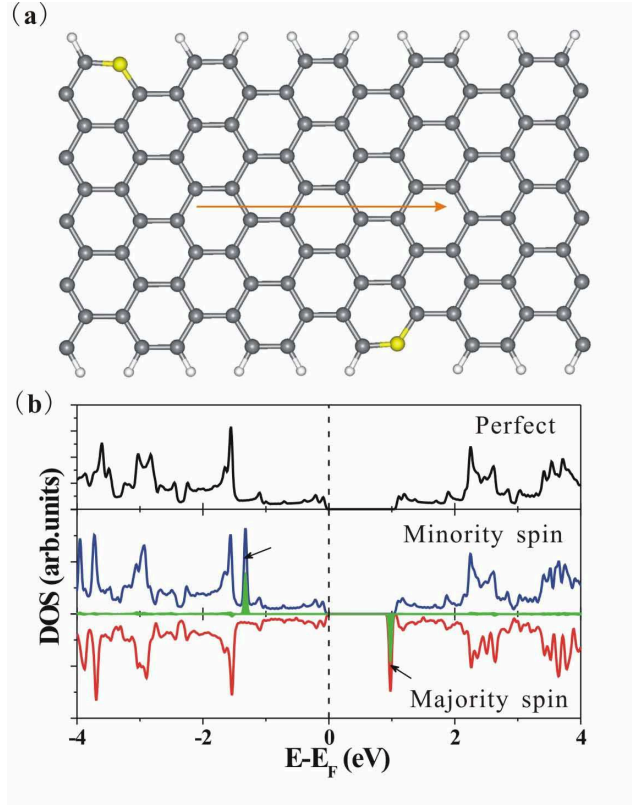}
\caption{(a) The optimized structure of 10-AGNR with edge dangling
bond (DB) defects, where the arrow shows the periodic direction.
The sites of DB defects are shown by yellow. C atoms and H atoms
are denoted by large and small (white) spheres, respectively. (b)
Density of states (DOS) of perfect 10-AGNR (top panel) and
spin-polarized DOS of 10-AGNR with DB defects (middle and bottom
panels). The Fermi level is set to zero (the top of valence band).
The middle (blue) and bottom (red) panels in the figure correspond
to minority spin and majority spin, respectively. The green area
corresponds to local DOS of the two carbon atoms (yellow balls in
(a)) with DB defects.}
\end{figure}

\section{Results and discussion}
DB defects would exist at both edges of GNRs due to the fact that
the two edges of a GNR are equivalent in nature. So we will focus
our study on double-edge-defect case (\emph{i. e.}, both edges
have DB defects). What is more, we also preform test calculations
for the single-edge-defect case (DB defects only exist at one
edge), and find little difference in essential results compared
with the double-edge-defect case. The structure with one DB defect
per edge per five unit cells in the ribbon axis direction is
adopted in our calculations, as shown in Fig. 1a, corresponding to
a DB defect concentration of 0.04 \AA$^{-1}$. [Several different
initial configurations with one DB defect per edge have been
considered and we find that the one shown in Fig. 1a is most
favorable (stable) in energy.] Due to the dangling $\sigma$-bonds
at the edges, the ground state of this system is spin-polarized
with the magnetic moment of 1 $\mu$B per DB, which is localized at
the carbon atoms with DB defects. Spin-polarized density of states
(DOS) of this system is shown in the middle and bottom panels of
Fig. 1b. For better comparison, the DOS of prefect 10-AGNR with
the same supercell is also presented (the top panel). It can be
seen that the perfect 10-AGNR is paramagnetic semiconductor with a
band gap of 1.2 eV, consistent with our previous
study\cite{Qimin}. When there are DB defects at the ribbon edges,
two new peaks appear in the DOS (indicated by the arrows in Fig.
1b). The local density of states (LDOS) analysis (the green area
in Fig. 1b) shows that the two peaks are mainly contributed by the
edge carbon atoms with DB defects. These two DB states are
localized, respectively, within the valence band for minority spin
and at the bottom of conduction band for majority spin, different
from the usual DB states which are localized within band
gap\cite{Chenchen}.

%fig02
\begin{figure*}[tbp]
\includegraphics[width=12.0cm]{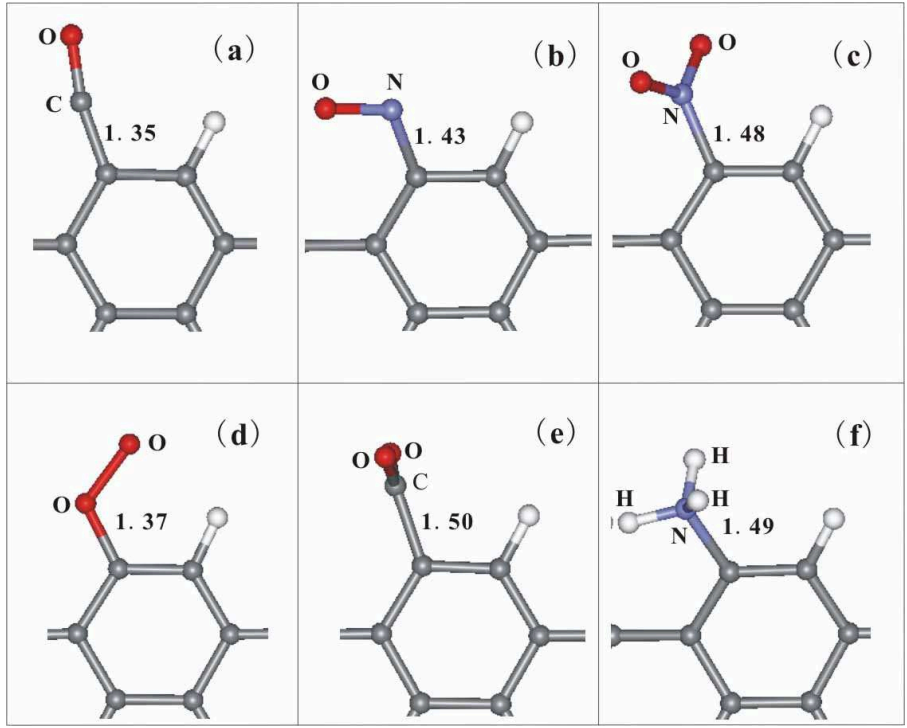}
\caption{Optimized structure of 10-AGNRs with gas molecule
adsorption: (a) CO, (b) NO, (c) NO$_{2}$, (d) O$_{2}$, (e)
CO$_{2}$, and (f) NH$_{3}$. We only show the structure around the
adsorbed molecule.}
\end{figure*}

\begin{table*}
\caption{\label{tab:table2} Calculated adsorption energies
($E_{\rm ads}$) and charge transfer ($\emph{CT}$) from the gas
molecules to the 10-AGNR.}
\begin{ruledtabular}
\begin{tabular}{cccccccc}
Molecules & CO & NO & NO$_{2}$ & O$_{2}$ & N$_{2}$ & CO$_{2}$ &
NH$_{3}$
 \\[2pt]
\hline
\ $E_{\rm ads}$ (eV) & -1.34  & -2.29 & -2.70 & -1.88 & 0.24 & -0.31 & -0.18 \\[2pt]
\ $\emph{CT}$ ($\emph{e}$) & -0.30  & -0.55 & -0.53 & -0.78 & / & -0.41 & 0.27 \\[2pt]
\end{tabular}
\end{ruledtabular}
\end{table*}

We start by investigating the adsorption geometries of seven gas
molecules on 10-AGNR with DB defects (Fig. 1a). Only the gas
molecules adsorption around the sites of DB defects is considered.
Several different initial orientations of gas molecules on
10-AGNRs are adopted in searching the most stable configurations.
Fig. 2 shows the top views of 10-AGNR with adsorbed molecules. We
find that different gas molecules prefer different geometries in
the adsorption: (a) CO molecule lies in the AGNR plane with C-C
distance of 1.35 \AA~ and a C-C-O angle of ~168.6$^\circ$. The
bond length of the adsorbed CO is 1.18 \AA, a little longer than
that of an isolated molecule (1.13 \AA), indicating that
adsorption process will weaken the original C-O bond of gas
molecule. (b) NO molecule sits out of the AGNR plane with a C-N-O
angle of ~118.2$^\circ$, and the C-N and N-O distances are 1.43
\AA~ and 1.24 \AA, respectively. Also, the bond length of the
adsorbed NO is 0.07 \AA~ longer than that of an isolated molecule.
(c) The adsorbed NO$_{2}$ has C-N and N-O distances of 1.47 \AA~
and 1.25 \AA~ and a O-N-O angle of 127$^\circ$, and the O-N-O
plane is tilted ~60$^\circ$ with respect to the ribbon plane. Our
results are consistent with previous studies\cite{Hao, Ferrari}.
(d) O$_{2}$ sits in the ribbon plane with C-O and O-O distances of
1.37 \AA~ and 1.38 \AA~ respectively and a C-O-O angle of
~116.2$^\circ$. The bond length of O$_{2}$ molecule increases by
0.15 \AA~ after adsorption. (e) The geometry of CO$_{2}$ adsorbed
on the ribbon edge is different from that of an isolated CO$_{2}$
molecule: it is not linear structure anymore but similar to the
NO$_{2}$ configuration. This indicates that the bond of CO$_{2}$
transfers from \emph{sp} into \emph{sp$^2$} hybridization in order
to lower the total energy. The C-C and C-O distances are 1.51 \AA~
and 1.26 \AA~ respectively, and the C-O-C angle is 127$^\circ$.
(f) The adsorbed NH$_{3}$ molecule sits 1.49 \AA~ away from the
edge carbon atom, the H-N distance is 1.03 \AA~ and the dihedral
angle is about 21$^\circ$ between the C-N bond and the AGNR plane.
Besides the above gas molecules, we also study the N$_{2}$
molecule adsorption on the AGNR, and find that it is difficult for
N$_{2}$ to adsorb on the ribbon due to its inert nature.

%fig03
\begin{figure*}[tbp]
\includegraphics[width=16.0cm]{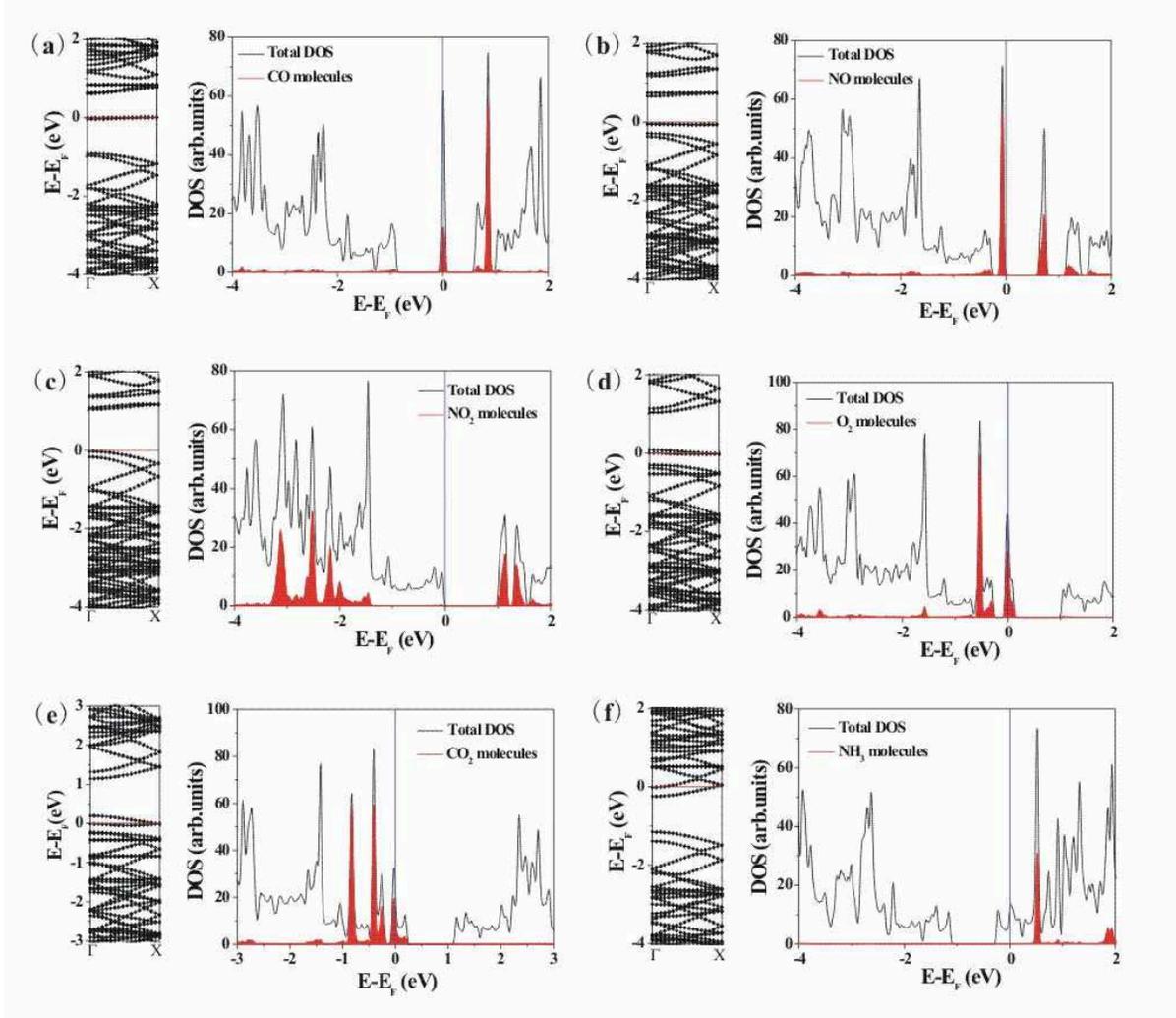}
\caption{Band structure and density of states (DOS) of 10-AGNRs
with gas molecule adsorption: (a) CO, (b) NO, (c) NO$_{2}$, (d)
O$_{2}$, (e) CO$_{2}$, and (f) NH$_{3}$. The LDOS of gas molecules
is also plotted (red filled area under DOS curve). The Fermi level
is set to zero.}
\end{figure*}

The calculated adsorption energies of the gas molecules with the
10-AGNRs are shown in Table 1. Herein, The adsorption energy is
defined as: $E_{\rm ads}  = \{ E_{\rm tot} ({\rm{ribbon}} +
m{\rm{Molecule}}) - E_{\rm tot} ({\rm{ribbon}}) - nE_{tot}
({\rm{Molecule}})\} /m$, where $E_{\rm tot} ({\rm{ribbon}} +
m{\rm{Molecule}})$, $E_{\rm tot} ({\rm{ribbon}})$, and $E_{\rm
tot} ({\rm{Molecule}})$ are the total energies of the AGNR with
molecule adsorption, the isolated AGNR (with DB defects) and the
molecules, respectively. And $m$ is the number of molecules
adsorption on the AGNR. The results reveal that all these
adsorption configurations are energetically favorable except that
the N$_{2}$ adsorption process is endothermic reaction (note: the
negative adsorption energy corresponds to the exothermic
reaction). The adsorption energies of CO, NO, NO$_{2}$, and
O$_{2}$ are all larger than 1 eV, corresponding to strong
chemisorption. The adsorption energies for the NH$_{3}$ and
CO$_{2}$ on AGNRs are -0.18 eV and -0.31 eV respectively,
indicating that the adsorption are between weak chemisorption and
strong physisorption. The above results illuminate that gas
molecule adsorption at AGNR edges is quite different from the weak
physisorption of gas molecules on the graphene
surface\cite{F.Schedin, O.Leenaerts}. Furthermore, the Bader
analysis\cite{Bader} of the charge distribution is used to
understand the nature of the interaction between the gas molecules
and the AGNRs, and to evaluate the induced effects on the
molecules. The trend of calculated charge transfer (Table 1) can
be understood on the basis of relative electron-withdrawing or
-donating capability of the adsorbed molecular groups. From Table
1 (note: the positive value means a charge transfer from the
adsorbed molecule to the AGNR), we can see that CO, NO, NO$_{2}$,
O$_{2}$, and CO$_{2}$ have electron-withdrawing capability, while
NH$_{3}$ is electron-donating functional molecule. It is well
known that NO$_{2}$ and O$_{2}$ are relatively strong
electron-withdrawing molecules while NH$_{3}$ is relatively strong
electron-donating molecule in other carbon-based
materials\cite{Kong, Collins, F.Schedin, O.Leenaerts, Andzelm}.
Comparing with gas adsorption on the graphene (GNR)
surface\cite{O.Leenaerts}, the values of the charge transfer are
much larger. This is consistent with our conclusion that the
interaction between the gas molecules and the GNR edges is much
stronger than that of surface.

%fig04
\begin{figure*}[tbp]
\includegraphics[width=14.0cm]{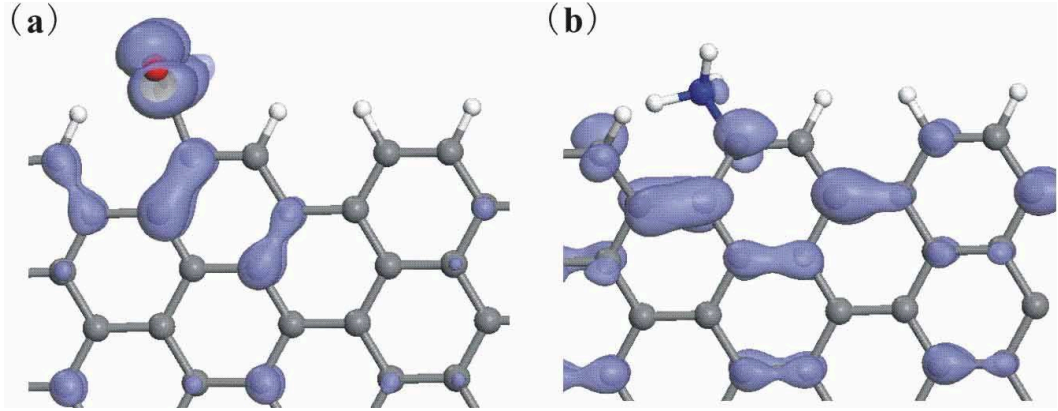}
\caption{Iso-surface plots of the partial charge density at
$\Gamma$ point of the band crossing the Fermi level for the
10-AGNRs with (a) CO$_{2}$ adsorption and (b) NH$_{3}$ adsorption.
The iso-value is 0.01 e/\AA$^{3}$.}
\end{figure*}

The calculated band structures and DOS of 10-AGNRs with molecule
adsorption are shown in Fig. 3. Comparing with the DOS of AGNR
(Fig. 1b), the total DOS of the system and LDOS of the molecules
show that these molecules modulate the electronic property of
AGNRs in different manners: i) CO and NO molecules adsorption
introduces impurity states in the band gap and the Fermi levels of
two systems cross these states, as shown in Figs. 3a and 3b.
Therefore, gas adsorption will decrease the original band gap, and
probably have some influence on the optical properties of AGNRs.
For CO adsorption, there are two half-occupied states in the band
gap, but we do not expect these impurity states can enhance the
conductance of the system because these states are very localized
and deep in the original band gap. It would be very difficult for
charge carriers to transit between the valence (or conduction)
band and impurity states at finite temperature. ii) LDOS analysis
(Fig. 3c) shows that NO$_{2}$ adsorption will introduce
fully-occupied states which are strongly hybridized with the
original ``bulk'' states in the valence band and these states are
nonlocalized. It suggests that the interaction between NO$_{2}$
molecules and dangling bonds of the ribbon is very strong,
consistent with the calculated adsorption energy. The Fermi level
is pinned in the top of valence band, which is the same as the
case of the ribbon without molecule adsorption, so the system is
still semiconducting. iii) Figs. 3d and 3e show the cases of
O$_{2}$ and CO$_{2}$ adsorptions, respectively. From LDOS analysis
we can see that the states contributed by CO$_{2}$ (or O$_{2}$)
molecules are localized around the top of valence band and
hybridize with the original valence band. Partial charge density
analysis (Fig. 4a) shows that the states near the Fermi level are
quite localized and mainly contributed by CO$_{2}$ molecule and
the carbon atoms of ribbon around the CO$_{2}$ molecule. This
suggests that the conductance of this system can not be enhanced
notably.  When the molecular doping concentration is low enough,
these impurity states will become more localized on the gas
molecules. But due to these half-occupied impurity states being
near the top of valence band, the electrons of the valence band
can transit into these states and the system will exhibit
\emph{p}-type semiconducting behavior at finite temperature. iv)
NH$_{3}$ molecule adsorption induces unoccupied local states in
the conduction band, and more importantly, the Fermi level is
shifted into original conduction bands, resulting in
$\emph{n}$-type semiconducting behavior (Fig. 3f). Furthermore,
partial charge density analysis (see Fig. 4b) shows that the
states near the Fermi level are mainly contributed by the carbon
atoms of the ribbon rather than NH$_{3}$ molecules. The above
results indicate a transition from semiconducting to conducting
behavior after NH$_{3}$ molecule adsorption. The stabilization of
the Fermi level in the conduction band by the impurity resonant
levels was also reported in semiconductors doped by Cr and
Tl\cite{pss448,prb3903}.

Among all gas molecules considered, obviously NH$_{3}$ molecule
adsorption can greatly enhance the conductance of AGNRs, where the
system will exhibit metallic behavior after NH$_{3}$ adsorption.
CO$_{2}$ and O$_{2}$, on the other hand, may enhance the
conductance of GNRs to some extent at finite temperature since the
CO$_{2}$ and O$_{2}$ adsorption will turn AGNRs to \emph{p}-type
semiconductor. Based on the analysis, we can expect that AGNRs may
act as effective sensor to detect NH$_{3}$ out of other gas
molecules discussed above by measuring the change of conductance
after the gas adsorption. What is more interesting, the adsorption
energy of NH$_{3}$ molecule is only -0.18 eV (Table I), so
NH$_{3}$ molecules can be desorbed at higher temperature. This
implies that GNR senors could be recycled more than once.

%fig05
\begin{figure}[tbp]
\includegraphics[width=8.5cm]{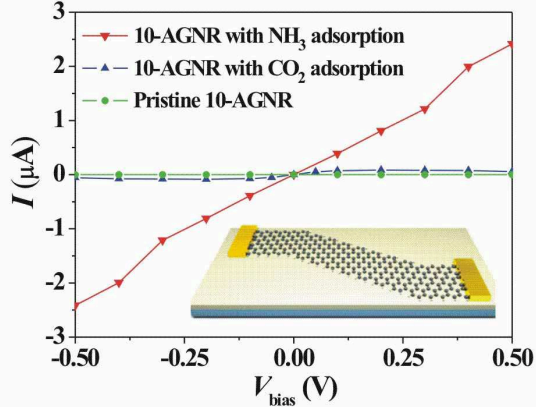}
\caption{The $I$-$V_{\rm bias}$ curves for the GNR sensor before and
after the adsorption of NH$_{3}$ and CO$_{2}$. The inset shows the
schematics of such a GNR sensor, consisting of one 10-AGNR
(detection region) and two metallic 7-ZGNRs leads. The gas molecules
can be adsorbed around the DB defects of the GNR sensor.}
\end{figure}

%A 8.60 nm long 10-AGNR with one DB defect per edge has been used as
%detection region and two semi-infinite metallic 7-ZGNRs have been
%used as leads for simplicity, as shown in the inset in the Fig. 5,
%and gas molecules can be adsorbed on the DB defect sites of such GNR
%junction.

A design of a GNR-based junction (GNR sensor) to detect NH$_{3}$
is given in the inset of Fig. 5 as an example. It contains a 8.60
nm long 10-AGNR with one DB defect per edge as the detection
region and two semi-infinite metallic 7-ZGNRs as the leads. Gas
molecules can be adsorbed on the DB defect sites, and the
conductance is measured by applying a bias voltage through the
junction. The DB defect concentration in this model is about
0.011/\AA~ per edge, which is practical in experiments. We
calculate a series of current versus bias voltage ($I$-$V_{\rm
bias}$) curves for such GNR junction with different gas molecule
adsorption on the edges. For the sake of clarity, we only show
$I-V_{\rm bias}$ curves for the AGNRs before and after NH$_{3}$
and CO$_{2}$ adsorption due to the fact that the currents induced
by other gas molecules adsorption are almost zero (much smaller
than the currents induced by CO$_{2}$ or NH$_{3}$ adsorption). As
shown in Fig. 5, without gas molecule adsorption, the channel (GNR
sensor) exhibits a semiconducting behavior, and the current is
always zero even under a bias of 0.5 V. After NH$_{3}$ adsorption,
however, the current increases notably and $I$-$V_{bias}$ curve is
nearly linear, corresponding to a metallic behavior of ohmic
contact. CO$_{2}$ adsorption can also increase the current, but
its value is much smaller than that induced by NH$_{3}$
adsorption. This phenomenon indicates that NH$_{3}$ can be
detected out of other gases by applying a bias voltage upon the
GNR junction, which is consistent with our analysis based on
electronic properties.

%Also, we can distinguish NH$_{3}$ molecule (\emph{n}-type dopant)
%and CO$_{2}$ (\emph{p}-type dopant)by applying gate
%voltages\cite{Qimin, Bing} in experiments.

\section{Summary}

In summary, we have performed first-principles calculation to
study the adsorption geometries and electronic structure of
graphene nanoribbons with gas molecule adsorption. We find that
NH$_{3}$ molecule adsorption can significantly influence the
electronic and transport properties of AGNRs, while other gas
molecules have little effect. Based on this characteristic, we
demonstrate that an AGNR can be used to detect NH$_{3}$ molecules
out of many familiar gas molecules. Furthermore, our work also
suggests an effective way to fabricate \emph{n}-type
(\emph{p}-type) transistors by NH$_{3}$ (CO$_{2}$ or O$_{2}$)
adsorption on graphene nanoribbons.

\acknowledgments This work was supported by the Ministry of
Science and Technology of China (Grant Nos. 2006CB605105 and
2006CB0L0601), the Natural Science Foundation of China (Grant Nos.
10674077 and 10774084) and the Ministry of Education of China.

\newpage

\end{document}